\pdfoutput=1
\documentclass[fleqn,usenatbib]{mnras}
\usepackage{newtxtext,newtxmath}
\usepackage[T1]{fontenc}
\usepackage{ae,aecompl}

\usepackage{graphicx}	
\usepackage{amsmath}	
\usepackage{amssymb}	
\usepackage{color}
\usepackage{relsize}
\usepackage{hyperref}
\hypersetup{pdfauthor={J. M. Corral-Santana},
               pdftitle={The long-term optical evolution of the black hole candidate MAXI J1659--152},
               pdfkeywords={X-rays binaries, black holes, MAXI J1659--152}}

\newcommand{\maxi}{MAXI\,J1659$-$152~}
\newcommand{\ha}{H$\alpha$~}
\newcommand{\diag}{($r'- {\rm H}\alpha$) vs ($r'-i'$)~}
\newcommand{\porb}{$P_{\rm orb}$~}
\newcommand{\ebv}{$E(B-V)$}
\newcommand{\xe}[2]{$#1\times 10^{#2}$}

\title[Long-term evolution of MAXI\,J1659$-$152]{The long-term optical evolution of the black hole candidate MAXI\,J1659$-$152}

\author[Corral-Santana et al.]{
Jes\'us~M. Corral-Santana$^{1,2}$\thanks{E-mail: jcorral@eso.org},
Manuel~A.~P. Torres$^{3,4,5}$, 
Tariq Shahbaz$^{3,4}$,
\newauthor
Elizabeth~S. Bartlett$^{1}$, 
David~M. Russell$^{6}$, 
Albert K.~H.~Kong$^{7}$, 
Jorge Casares$^{3,4,8}$,
\newauthor 
Teodoro Mu\~noz-Darias$^{3,4}$, 
Franz~E. Bauer$^{2,9,10}$,
Jeroen Homan$^{11,5}$, 
Peter~G. Jonker$^{5,12}$, 
\newauthor 
Daniel Mata S\'anchez$^{3,4}$,
Thomas Wevers$^{12}$,
Pablo Rodr\'iguez-Gil$^{3,4}$,
Fraser Lewis$^{13,14}$
\newauthor 
and Laurien Schreuder$^{15}$
\\
$^{1}$European Southern Observatory (ESO), Alonso de C\'ordova 3107, Vitacura, Casilla 19, Santiago, Chile\\
$^{2}$Pontificia Universidad Cat\'olica de Chile, Vicu\~na-Mackenna 4860, 
	Macul, Santiago, Chile\\
$^{3}$Instituto de Astrof\'isica de Canarias, V\'ia L\'actea s/n, E-38205, 
	La Laguna, Spain\\
$^{4}$Universidad de La Laguna, Dpto. de Astrof\'isica, Astrof\'isico 
	Francisco S\'anchez s/n, E-38206, La Laguna, Spain\\
$^{5}$SRON, Netherlands Institute for Space Research, Sorbonnelaan 2, 3584 CA Utrecht, The Netherlands\\
$^{6}$New York University Abu Dhabi, PO Box 129811, Abu Dhabi, UAE\\
$^{7}$National Tsing Hua University, Department of Physics and Institute of Astronomy, No. 101 Sect. 2 Kuang-Fu Road, 30013, Hsinchu, Taiwan\\
$^{8}$Department of Physics, Astrophysics, University of Oxford, Keble Road, Oxford OX1 3RH, UK\\	
$^{9}$Millenium Institute of Astrophysics (MAS), Nuncio Monse\~nor S\'otero Sanz 100, Providencia, Santiago, Chile\\
$^{10}$Space Science Institute, 4750 Walnut Street, Suite 205, Boulder, Colorado 80301\\
$^{11}$Eureka Scientific, Inc., 2452 Delmer Street, Oakland, California 94602\\
$^{12}$Department of Astrophysics/IMAPP, Radboud University Nijmegen, P.O. box 9010, 6500 GL Nijmegen, The Netherlands\\
$^{13}$Faulkes Telescope Project, School of Physics, and Astronomy, Cardiff University, The Parade, Cardiff CF24 3AA, UK\\
$^{14}$Astrophysics Research Institute, Liverpool John Moores University, 146 Brownlow Hill, Liverpool L3 5RF, UK\\
$^{15}$Astronomical Institute Anton Pannekoek, University of Amsterdam, P.O. Box 94249, 1090 GE Amsterdam, The Netherlands\\
}

\date{Accepted 2017 December 4. Received 2017 December 1; in original form 2007 October 25}

\pubyear{2017}
\volume{{\rm in press}}
\begin{document}
\label{firstpage}
\pagerange{\pageref{firstpage}--\pageref{lastpage}}
\maketitle

\begin{abstract}
\label{sec:abs}
We present 5 years of optical and infrared data of the black hole 
candidate \maxi covering its 2010 outburst, decay and quiescence. 
Combining optical data taken during the 
outburst decay, we obtain an orbital period of $2.414\pm0.005$\,h, 
in perfect agreement with the value previously measured from X-ray dips. 
In addition, we detect a clear \ha excess in \maxi with data taken 
during the outburst decay. 
We also detect a single hump modulation 
most likely produced by irradiation. Assuming that the maximum 
occurs at orbital phase 0.5, we constrain the phase of the X-ray 
dips to be $\sim0.65$.
We also detect the quiescent optical counterpart at $r'=24.20\pm0.08$, 
$I=23.32\pm0.02$ and $H=20.7\pm0.1$. These magnitudes provide colour 
indices implying an M2--M5 donor star assuming 60\% contribution from a 
disc component in the $r'$-band. 
\end{abstract}




\section{Introduction}
\label{sec:intro}
X-ray binaries are systems formed by a neutron star or a black hole 
(BH) accompanied by a star which is feeding the compact source via an 
accretion disc. The mass of the companion star determines the 
main mode of accretion, e.g. whether produced through stellar winds 
in systems with massive companions or through Roche lobe overflow 
in systems with cool stars. Most of the systems harbouring a BH have 
been found in X-ray transients (XRTs), a type of X-ray binary with 
sporadic outburst episodes 
--caused by thermal-viscous instabilities in the accretion disc 
\citep[see e.g. ][]{Lasota2001,King1996}-- followed by an e-folding 
decay towards the quiescent state, where the system can reside for decades 
to centuries.
During outburst, XRTs increase their brightness by several orders of 
magnitude, such that they can be detected by the all-sky monitors 
onboard X-ray satellites.

\maxi (initially referred to as GRB\,100925A) was detected during 
an outburst on 2010 September 25 simultaneously by the Gas Slit Camera 
(GSC) mounted on the Monitor of All-sky X-ray Image (\textit{MAXI}) 
\citep{Negoro2010} and the Burst Alert 
Telescope (BAT) mounted on the \textit{Swift} telescope 
\citep{Mangano2010}. \cite{Ugarte2010} found broad emission lines from 
the Balmer series as well as He\,\textsc{ii} and Ca\,\textsc{ii}, 
confirming the Galactic 
origin and the X-ray binary nature. It was classified as a BH 
candidate based on its X-ray spectral and timing properties 
\citep{Kalamkar2010,Kalamkar2011,Munoz-Darias2011}, 
which resemble those typically found in confirmed BH XRTs.
(see, e.g., \citealt{Belloni2011,Belloni2016}). 
A radio counterpart was also 
detected with the Westerbork Synthesis Radio Telescope showing 23\% 
linear polarization \citep{vanderHorst2010} and the European Very Large 
Base Interferometer (eVLBI) network. The coordinates of the radio source 
are $\alpha$ (J2000) = 16h59m01.676891s and $\delta$ (J2000) = 
-15\degr 15\arcmin 28.73237\arcsec, with 690 and 220\,$\mu$as 
uncertainties, respectively \citep{Paragi2013}.

The optical counterpart was promptly detected by \textit{Swift}/UVOT 
\citep{Marshall2010} and several ground based telescopes 
\citep[e.g.][]{Jelinek2010,DeCia2010}.
By analysing pre-outburst survey images, 
\cite{Kong2010} reported an optical
counterpart consistent with the coordinates of the X-ray source. 
Later, \cite{Kong2012} reported a quiescent $r'$-band magnitude in the 
range 23.6$-$23.8 from postoutburst images. 
However, \cite{Kuulkers2013} proposed that the quiescent counterpart found 
by \cite{Kong2010} and \cite{Kong2012} was too bright to be the true 
counterpart and probably was a foreground star in our line-of-sight. 
To conclude this, they used the \cite{Shahbaz1998} relation corrected 
for inclination effects \citep{Warner1987,Miller-Jones2011c} and 
proposed a magnitude for the quiescent counterpart in the range 
$V=26.2-27.5$.

Dips in the X-ray flux were detected during outburst 
with \textit{Swift/X-ray telescope} \citep{Kennea2010,Kennea2011}, 
\textit{INTEGRAL} 
\citep[International Gamma Ray Astrophysics Laboratory;][]{Kuulkers2010a} 
and \textit{XMM-Newton} \citep{Kuulkers2010}, 
recurring approximately every 2.4$-$2.5\,h 
\citep{Kuulkers2010,Kuulkers2013}. 
 This allowed \cite{Kuulkers2013} to obtain an accurate orbital period 
 ($P_{\rm orb}$) of $2.414\pm0.005$\,h, being the shortest \porb 
 reported for a BH candidate hitherto \citep[see, e.g., BlackCAT
\footnote{www.astro.puc.cl/BlackCAT};][]{Corral-Santana2016}. 
 \cite{Kuulkers2013} constrained the inclination to be between 
 65\degr~and 80\degr~due to the lack of eclipses, and the presence of 
 the aforementioned periodic absorption dips that obscured up to 90\% 
 of the total emission. 

The distance of \maxi is not well constrained. With the reported 
quiescent counterpart, \cite{Kong2012} assumed an M2 or M5 dwarf 
companion which contributes 50\% to the total optical flux 
\citep{Jonker2012a} to derive a distance of $\sim5.5$\,kpc and 
$\sim3$\,kpc, respectively. \cite{Kennea2011} estimated a distance 
of $>6.1$\,kpc from the properties of the X-ray emission whereas, 
\cite{Kaur2012} obtained a distance of $4\pm1$\,kpc from the interstellar 
lines of the optical spectra. \cite{Jonker2012a} derived a distance of 
$6\pm2$\,kpc assuming an M2\,V companion star with a 50\% 
contribution of the accretion disc to the $R$-band magnitude.
Finally, using the \cite{Shahbaz1998} relation together with the 
estimated Galactic hydrogen column density ($N_H$) given by 
\citet{Kalberla2005}, \cite{Kuulkers2013} obtained a distance to \maxi 
of $8.6\pm3.7$\,kpc. However, this relation does not take into account 
inclination effects, yielding overestimated values 
\citep{Miller-Jones2011c}. Thus, the distance would drop  
to $7.1\pm3.0$\,kpc for $i=65$\degr~or $4.0\pm1.7$\,kpc 
for $i=80$\degr~\citep{Kuulkers2013}.

Interestingly, there is a growing population of XRTs with short \porb 
located at high Galactic latitudes. \maxi has a height above the 
Galactic plane of $z=2.4\pm1.0$\,kpc. 
This can be compared to 
Swift\,J1753.5$-$0127 with $z=1.1$\,kpc and \porb$=3.24$\,h 
\citep{CadolleBel2007,Zurita2008}; 
XTE\,J1118+480 with $z=1.6$\,kpc and \porb$=4.08$\,h 
\citep{Zurita2002,Torres2004,Gelino2006}
and more recently Swift\,J1357.2$-$0933 with $z>1.75$\,kpc and \porb$=2.8$\,h 
\citep{MataSanchez2015,Corral-Santana2013,Shahbaz2013}. It has been 
proposed by \cite{Yamaoka2012} that \maxi is probably a runaway BH XRTs 
kicked out from the Galactic plane at a high speed similar to 
XTE\,J1118+480 \citep{Mirabel2001}. Although this scenario has 
not been established for Swift\,J1753.5$-$0127 and 
Swift\,J1357.2$-$0933 yet, the hypothesis is well supported by the 
properties found in \maxi \citep{Kuulkers2013}.

\section{Observations and data reduction}
\label{sec:obs}
\begin{figure}
	\includegraphics[width=\columnwidth]{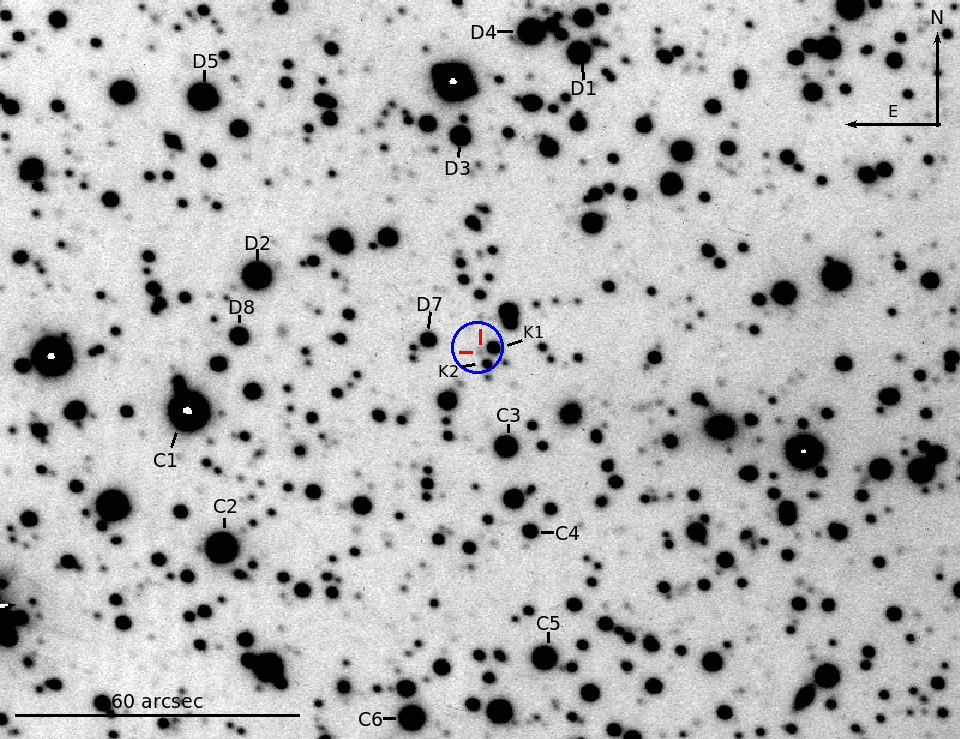}
    \caption{$I$-band finding chart of the field containing \maxi in 
    quiescence obtained on 2015 June 12 with VLT. The comparison stars 
    are labelled in black whereas \maxi is marked in red. The ID names 
    of the comparison stars correspond to those in 
    Table~\ref{tab:compar}. The two comparison stars used by 
    \citet{Kong2012} are labelled K1 and K2. The blue circle 
    represent the \textit{Swift/X-ray Telescope} positional error circle.}
    \label{fig:fc}
\end{figure}

In what follows, 
we detail the observations and reduction of the data taken with the 
Gran Telescopio Canarias (GTC), Gemini South (GS), Very Large 
Telescope (VLT), William Herschel (WHT), Isaac Newton (INT), Faulkes 
North and South (FTN and FTS, respectively) and  Mercator (MER) 
telescopes. A log of the observations is shown in 
Table~\ref{tab:log}. In Figure~\ref{fig:fc} we present a finding 
chart with the comparison stars used for the photometry, the 
magnitudes of which are listed in Table~\ref{tab:compar}.

\begin{table}
    \small
	\centering
	\caption{Log of the photometry. Each column represents the date 
	of observation, telescope, detector, filter, exposure time and 
	total number of images, respectively.
	}
	\label{tab:log}
	\begin{tabular}{cccccc} 
	    \hline
		UT date & Telescope$^*$ & Detector & Filter & Exp. (s) & No. \\
		\hline
		2010-09-26 & FTN & Merope   & $i'$ & 300 & 5  \\
		2010-09-28 & FTS & Merope   & $i'$ & 100 & 14 \\
		2010-10-05 & FTS & Merope   & $i'$ &  40 & 11 \\
		2011-05-21 & FTS & Spectral & $i'$ & 200 & 1 \\
		2011-05-25 & FTS & Spectral & $i'$ & 200 & 1  \\
		2011-05-26 & FTS & Spectral & $i'$ & 200 & 1  \\
		2011-05-27 & FTS & Spectral & $i'$ & 200 & 1  \\
 		2011-05-30 & Mer & Merope   & $R$ & 120 & 117 \\
		2011-05-31 & Mer & Merope   & $R$ & 120 & 132 \\
		2011-06-02 & FTN & Spectral & $i'$ & 200 & 24 \\
		2011-06-03 & FTN & Spectral & $i'$ & 200 & 4  \\
		2011-06-06 & FTN/S&Spectral & $i'$ & 200 & 37 \\
		2011-06-28 & WHT & ACAM     & $r'$ & 240 & 42 \\
		2011-07-27 & FTS & Spectral & $i'$ & 200 & 11 \\
		2011-07-31 & INT & WFC      & $r'$ & 120 & 2  \\     	
     	2011-08-03 & FTS & Spectral & $i'$ & 200 & 1  \\
		2011-08-26 & FTS & Spectral & $i'$ & 200 & 1  \\
		2011-08-28 & WHT & ACAM     & $r'$ & 200 & 2\\
    	     "     &  "  &   "      & $i'$ & 150 & 2\\
		     "     &  "  &   "      & H$\alpha$ & 600 & 3\\
		2012-06-08 & INT & WFC      & $r'$ & 1800 & 4 \\
		2013-04-13 & GS  & GMOS     & $i'$ & 603 & 13 \\
		2013-05-04 & GS  & GMOS     & $i'$ & 603 & 12 \\
        2013-06-06 & VLT & FORS2 & $I$ & 90+120 & 2\\
		2014-02-26 & WHT & LIRIS & $H$ & 30 & 180 \\
		2014-04-09 &  "  &   "   & $J$ & 60 & 90\\
		2014-06-30 & GTC & Osiris & $r'$ & 180 & 4 \\
		2014-07-20 & WHT & LIRIS & $H$ & 30 & 120\\		
		2014-09-01 &  "  &   "   & $K$ & 15 & 278\\		
		2015-06-12 & VLT & FORS2 & $I$ & 240 & 11 \\ 
		\hline
		\multicolumn{6}{l}{$^*$ See text for the definition of the acronyms}
	\end{tabular}
\end{table}

\subsection{Photometry from Australia and Hawaii}
\label{sec:ft} 
Optical images of MAXI\,J1659-152 were taken with the 2-m FTN and FTS 
telescopes situated in Haleakala, Hawaii (USA) and Siding Spring 
(Australia), respectively. The observations are part of an ongoing 
monitoring campaign of $\sim 40$ low-mass X-ray binaries 
\citep{Lewis2008}.
The observations mostly covered the outburst and decay phases, since during 
quiescence the system is too faint to obtain clear detections with these 
telescopes. Useful images were taken in $i'$-band between 2010 September 
26 and 2011 August 26. We adjusted the exposure time to account for 
the change of brightness of the system, between 40 
and 300\,s. The observations obtained in 2010 were taken with the Merope 
detector which had a pixel scale of 0.278\,\arcsec~pix$^{-1}$ (binning 
$2\times2$), covering a field of view (FoV) of 
$4.7\,\arcmin\times4.7\,\arcmin$. Since 2011, the Spectral camera 
replaced Merope, providing a pixel scale of 0.304\,\arcsec~pix$^{-1}$~and a 
$10.5\,\arcmin\times10.5\,\arcmin$ FoV. The average image 
quality of the entire dataset is 1.3\,\arcsec.

Science images were de-biased and flat-fielded with an automatic 
pipeline. We then performed aperture photometry using 
\textsc{PHOT} in IRAF with a fixed aperture of 6.0 pixels 
($1.7$ and $1.8$\arcsec~for each camera, respectively) in the 
target and local comparison stars. 

\subsection{Observations from the Canary Islands}
\label{sec:others} 
We obtained images with several telescopes located at the Roque de los 
Muchachos Observatory on the island of La Palma, Spain. We observed 
with the 1.2-m MER telescope on 2011 May 30 and 31 using the Merope 
camera ($6.5\,\arcmin\times6.5\,\arcmin$ FoV, 
0.19\arcsec~pix$^{-1}$~plate scale) and the $R$-band filter. 
The exposure times were 120\,s for both nights, obtaining a total 
of 117 and 132 images, respectively, with a seeing better than 
0.9\arcsec~in all cases. 
The conditions during 2011 May 30 were degraded due to the 
presence of some cirrus during part of the night. 
Thus, data taken between UT\,01:44 and UT\,02:25 were 
removed due to its lack of quality to the detect the target.
The images were de-biased and flat-fielded following 
standard procedures in \textsc{IRAF}. Then, we applied differential 
optimal-aperture photometry \citep{Naylor1998} to obtain the magnitudes 
of local comparison stars and the object with tasks in \textsc{IRAF} 
and \textsc{IDL}.\\
We also observed MAXI\,J1659-152 with the 4.2-m WHT on 2011 June 28 
and August 28 using the Auxiliary-port CAMera (ACAM; \citealt{Benn2008}). 
This instrument provides a FoV of $8\,\arcmin \times 8\,\arcmin$~with 
a plate scale of 0.25\,\arcsec~pix$^{-1}$. The images were taken with 
the $r'-$, $i'-$ and H$\alpha-$band ($\lambda_{\rm c}=6553$, 
$\Delta\lambda=63$\,\AA) filters, with the latter only used on 2011 
August 28. The average seeing in each band along the entire 
FoV was 1.2, 1.2 and 1.1\arcsec, respectively. The reduction and 
analysis of these images were done as for the data taken with MER.

$J$, $H$ and $K_s$-band images were obtained with the Long-Slit 
Intermediate Resolution Infrared Spectrograph (LIRIS) on four nights 
between 2014 February 26 and September 01. This camera provides 
$4.27\arcmin\times4.27\arcmin$ FoV with a plate scale of 
0.25\arcsec~pix$^{-1}$. 
The observing conditions were variable with seeing ranging from 0.7 to 
2\,\arcsec. The exposure times used ranged from 15 to 60\,s depending 
on the filter (see Table~\ref{tab:log}). The standard data reduction 
was performed using the \textsc{lirisdr} package 
\citep{Alves2011}. Optimal aperture photometry 
\citep{Naylor1998} was carried out on the resulting combined 
images for each band to obtain the counts for \maxi and local 
comparison stars as well as a 2MASS star in the field 
(2MASS16590790$-$1515289). 
Differential photometry of \maxi with respect to the 
2MASS local standard was then performed, which also 
allow an absolute photometric calibration.

We also obtained images with the 2.5-m INT on 2011 July 31 and 2012 
June 08 and 09. In these campaigns we used the Wide Field Camera (WFC) 
mounted on the prime focus of the telescope with the $r'$-band filter, 
obtaining an average seeing of 2.0, 1.4 and 1.6\arcsec, 
respectively for each night. Given the large FoV of 0.3 square 
degrees of this camera, we reduced the exposure area 
in order to decrease the readout time. The images were 
de-biased and flat-fielded with \textsc{IRAF}.
We used optimal aperture photometry on the 2011 campaign. 
Given the faintness of the system in 2012 and the presence of nearby 
stars of similar brightness, we decided to perform point 
spread function (PSF) photometry using \textsc{DAOPHOT II/ALLSTAR} 
\citep{Stetson1987}. The PSF model was chosen interactively among the 
six different functions available in the software to fit the profile 
of point-like objects. This is repeated iteratively to reject 
non-stellar objects. Once the model converged, it is applied to all 
the detections in the field. On 2012 June 09 the 
weather and seeing conditions prevented us from detecting the system.

Photometric frames were taken on 2014 June 30 with the 10.4-m GTC 
telescope. Four 180\,s $r'$-band exposures were taken with the 
Optical System for Imaging and low-Intermediate-Resolution Integrated 
Spectroscopy (OSIRIS) with an average seeing of 0.8\arcsec. 
This instrument covers a $7.8\arcmin\times8.5\arcmin$ FoV in imaging 
mode with a plate scale of 0.254\arcsec~pix$^{-1}$~(binning $2x2$). 
As before, the images were de-biased and flat-fielded following standard 
procedures in \textsc{IRAF}. The reduced images were aligned 
and combined in order to detect the faint quiescent counterpart to 
\maxi with a higher signal-to-noise ratio. We then obtained PSF 
photometry on both the average and individual frames.

\subsection{Photometry from Chile}
\label{sec:vlt} 
We obtained $i'$-band images with the Gemini Multi-Object 
Spectrograph (GMOS) mounted on the 8-m GS situated in Cerro Pach\'on 
on 2013 April 13 and May 04. The observations were performed with 
the EEV detectors which had a pixel scale of 0.073\arcsec~pix$^{-1}$ 
covering a FoV of $5.5\arcmin \times 5.5\arcmin$ and a seeing 
better than 0.7\arcsec.

We also obtained $I$-band images with the 8-m VLT situated on Cerro 
Paranal on 2013 June 06 and 2015 June 12. We used the FOcal 
Reducer and low dispersion Spectrograph (FORS2) which provides a FoV 
of $6.8\arcmin \times 6.8\arcmin$~and a scale of 
0.25\arcsec~pix$^{-1}$ (binning $2x2$). The 2015 frames were taken in 
imaging mode and standard resolution collimator whereas the images taken 
in 2013 correspond to the acquisition frames of long-slit spectroscopy 
(see Torres et al., in preparation for more details).

We again reduced the GS and VLT images in the standard way with tasks 
on \textsc{IRAF} and performed PSF photometry in both the single and 
the average images. The source is only detected in six 
of the 11 VLT frames obtained in 2015 due to the combination of poor  
seeing (1.3\arcsec on average) and high airmass (>1.5). 
The 2013 data had 0.7\arcsec seeing.\\

\subsection{Calibration of the data}
\label{sec:calib}
Table~\ref{tab:log} lists all the data obtained in different bands 
and photometric systems. In order to calibrate the data, we searched  
for calibrated stars in our field in different public surveys. 
In our search we only found the star C1 [$\alpha(J2000)=16:59:05.900$, 
$\delta(J2000)=-15:15:41.74$] in the AAVSO Photometric All-Sky 
Survey (APASS, \citealt{Henden2009}) with magnitudes $r'=15.8\pm0.1$ 
and $i'=15.21\pm0.06$. Given the large uncertainty on the $r'$-band 
magnitude, we decided to calibrate this band with the check stars 
published by \citet{Kong2012} in his so called \textit{epoch 1} 
(hereafter labelled as K1 and K2), with derived $r'$-band magnitudes 
$20.738\pm0.004$ and $21.85\pm0.01$, respectively. We 
calibrated the $r'$-band magnitudes of the comparison stars listed in 
Table~\ref{tab:compar}, by obtaining a weighted average offset between 
our instrumental and the calibrated magnitudes. Thus, we derivate  
$r'=16.065\pm 0.005$ for C1. We adopted the C1 $i'$-band value given 
in APASS survey since \cite{Kong2012} did not obtain images in this band. 
The magnitudes of all the local standards in both bands are listed in 
Table~\ref{tab:compar}. We then derived the equivalent $R$- and 
$I$-band magnitudes of all the local comparisons applying the 
transformation equations given by \cite{Jordi2006}. 

\begin{table}
	\centering
	\caption{Magnitudes of all the comparison stars selected in the 
	field. Stars C1$-$C6 and D1$-$D8 were selected during outburst 
	and decay, respectively.	
	See Section~\ref{sec:calib} for details on the 
	photometric calibration.
	}
	\label{tab:compar}
	\begin{tabular}{lcc} 
		\hline
		ID & $i'$ & $r'$\\
		\hline
		C1 & $15.21\pm0.06$ & $16.065\pm0.005$ \\
		C2 & $16.31\pm0.06$ & $17.069\pm0.006$ \\
		C3 & $17.86\pm0.06$ & $18.795\pm0.006$ \\
		C4 & $19.54\pm0.06$ & $20.240\pm0.006$ \\
		C5 & $17.54\pm0.06$ & $18.256\pm0.008$ \\
		C6 & $17.35\pm0.06$ & $18.332\pm0.006$ \\ 
		D1 & $17.72\pm0.06$ & $18.450\pm0.006$ \\ 
		D2 & $16.83\pm0.06$ & $17.629\pm0.006$ \\
		D3 & $18.27\pm0.06$ & $19.023\pm0.006$ \\ 
		D4 & $17.13\pm0.06$ & $17.877\pm0.006$ \\ 
		D5 & $16.82\pm0.06$ & $17.532\pm0.006$ \\ 
		D7 & $19.35\pm0.06$ & $20.085\pm0.006$ \\ 
		D8 & $18.66\pm0.06$ & $19.407\pm0.006$ \\  
		\hline
	\end{tabular}
\end{table}
\begin{figure*}
	\includegraphics[width=0.8\textwidth,height=0.42\textheight]{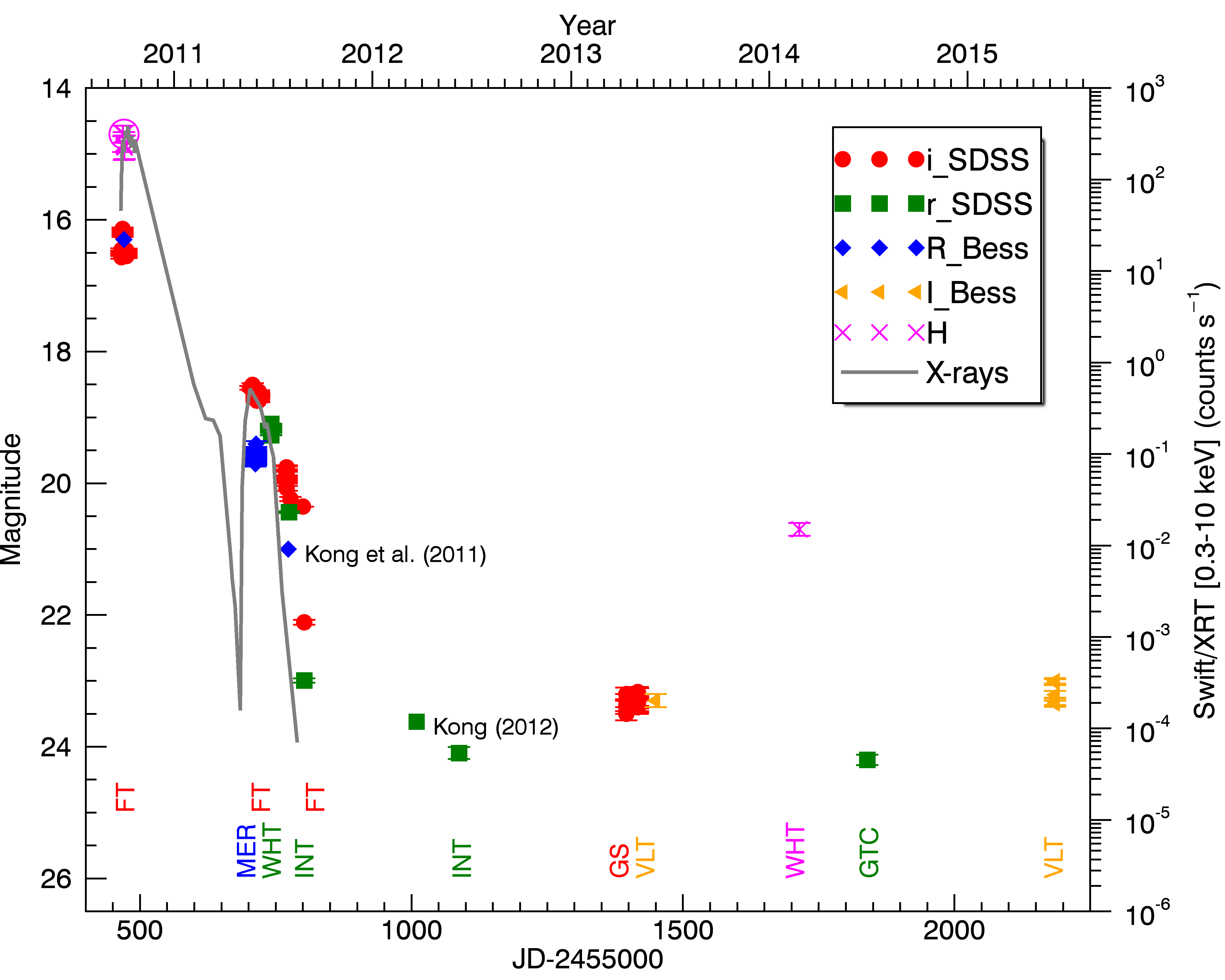}
    \caption{ 
    Optical and near-infrared light curve of \maxi. Each coloured 
    symbol represents a different band: $i'$ (red circles), $r'$ 
    (green squares), $R$ (blue diamonds), $I$ (orange left-pointing 
    triangles) and $H$-band (magenta X). We include $H$-band 
    photometry by \citet{vanderHorst2013} and 
    \citet[][encircled]{DAvanzo2010} and the $R$-band photometry by 
    \citet{Gorbovskoy2010} and \citet[][encircled]{DAvanzo2010}. 
    In addition, we added the $R$ and $r'$-band magnitudes given by 
    \citet{Kong2011} and \citet{Kong2012}, respectively. The solid 
    gray line represents the X-ray light curve including the 
    the reflare observed at $\sim$JD\,2455700 
    \citep[see Figure~4 in][]{Homan2013}.}
    \label{fig:evol}
\end{figure*}

\section{Results}
\label{sec:res}
In Figure~\ref{fig:evol} we show the light curve of \maxi covering  
the evolution since its 2010 outburst until 2015. 
It contains all the photometric data listed in Table~\ref{tab:log} 
together with the photometry published by 
\citet{DAvanzo2010}, \citet{Gorbovskoy2010}, \cite{Kong2011} 
and \citet{Kong2012}. In 
the bottom part of the figure, we added the acronyms of the telescopes 
used to obtain the data, using the same colour code as the photometric
points. 
Our images taken with MER on 2011 May 30-31 (JD\,2455714-5) 
were close to the peak of the X-ray rebrightening reported 
by \citet{Homan2013} ($\sim$JD\,2455700).

\subsection{\ha emission in outburst}
\label{sec:diag}
On 2011 August 28 we obtained, quasi-simultaneous (45\,min) 
$r'$, $i'$ and \ha images of the field containing \maxi with the WHT. 
The \ha emission in XRTs is produced in the accretion disc, 
which dominates the optical spectrum during the outburst in nearly 
all X-ray binaries. 

\begin{figure}
	\includegraphics[width=\columnwidth]{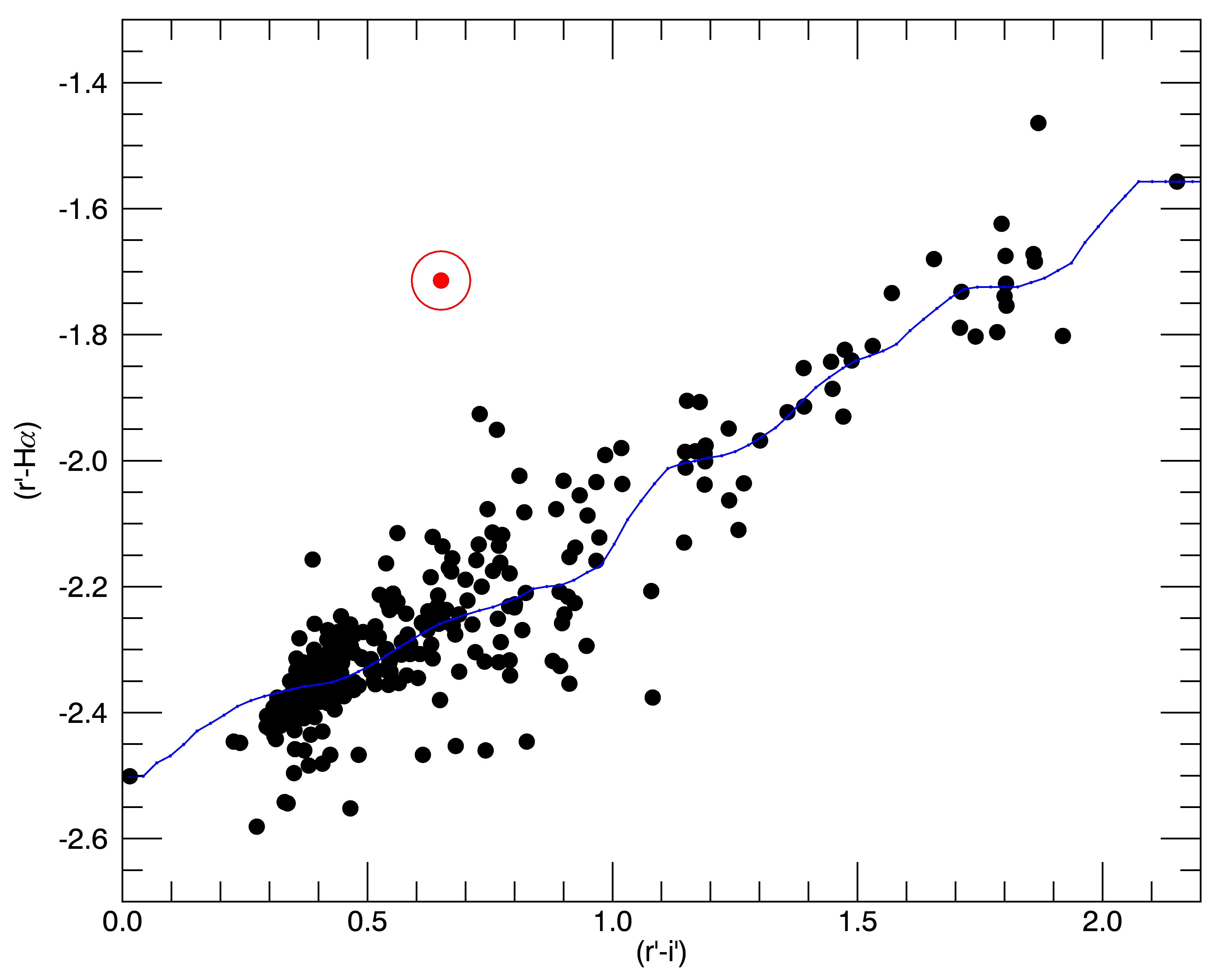}
    \caption{Uncalibrated \diag diagram of the stars in a field 
    of $3\arcmin \times 3\arcmin$ containing \maxi. The 
    main-sequence track is formed by the concentration of black circles, 
    with the reddening direction increasing towards higher ($r'-i'$) 
    and ($r'- {\rm H}\alpha$), whereas \maxi (encircled in red) is clearly 
    above them. This indicates that the optical counterpart detected 
    in our images shows the largest unexpected \ha excess, confirming it is the 
    true optical counterpart. }
    \label{fig:diag}
\end{figure}

With the instrumental magnitudes obtained in the above bands, we 
examine an \diag diagram. This has been used not only as a 
very efficient way to distinguish among different populations in 
the Galactic plane (see e.g. \citealt{Drew2005,Corradi2008,Corradi2010,Wevers2016})
but also to identify the missing optical counterpart of BH candidates 
(\citealt{Zurita2015}). Figure~\ref{fig:diag} shows the 
uncalibrated \diag diagram of all the stars observed in the detector 
FoV and detected in all three filters. 
The blue solid line is calculated as the running median of the 
($r'- {\rm H}\alpha$) colour of nearly all the observed stars in the field, 
providing a reference ($r'- {\rm H}\alpha$)$_0$ colour such that all 
objects with values above it will have an \ha excess.
The optical counterpart of \maxi (encircled in red) shows a clear 
\ha excess. 

From the colour-colour diagram (Figure~\ref{fig:diag}), it is 
possible to estimate the equivalent width of the \ha line [EW(H$\alpha$)]. 
To do so, we first need to obtain the \ha excess emission 
($\Delta$H$\alpha$) defined as the difference between the observed 
($r'- {\rm H}\alpha$)$_{\rm obs}$ colour of a star, and the reference 
value ($r'- {\rm H}\alpha$)$_0$ at the same ($r'- i'$)$_0$ colour. Then, 
the equivalent width EW(H$\alpha$) can be derived from equation 4 in 
\citet{DeMarchi2010}:
\begin{equation}
 {\rm EW(H\alpha)\sim RW} \times [1-10^{-0.4\times \Delta {\rm H \alpha}}]
\end{equation} 
where RW is the rectangular width of the filter in \AA~(see e.g. 
\citealt{DeMarchi2010} and \citealt{Beccari2014} for more details). 
To derive RW, we normalized to unity and resampled the \ha filter 
profile to 1\,\AA~and measured its equivalent width, obtaining a value 
of $38.63$\,\AA. Thus, with the equation above, we obtained 
EW(H$\alpha$)$\sim15$\,\AA~for MAXI\,J1659$-$152.

\subsection{Time series analysis}
\label{sec:period}
To search for a periodic modulation we performed a Lomb-Scargle 
analysis (\citealt{Lomb1976,Scargle1982}) on the data acquired on 2011 
May 30 and 31, June 02, 06 and 28. These nights were chosen as they have 
the best time coverage over the shortest baseline. Differential 
photometry of \maxi was performed with C4 and then detrended by 
subtracting the mean value of each night. Analysis was performed using 
the \textsf{astropy.stats.LombScargle} package.

\begin{figure*}
 \centering
 \includegraphics[width=0.8\textwidth,height=0.42\textheight]{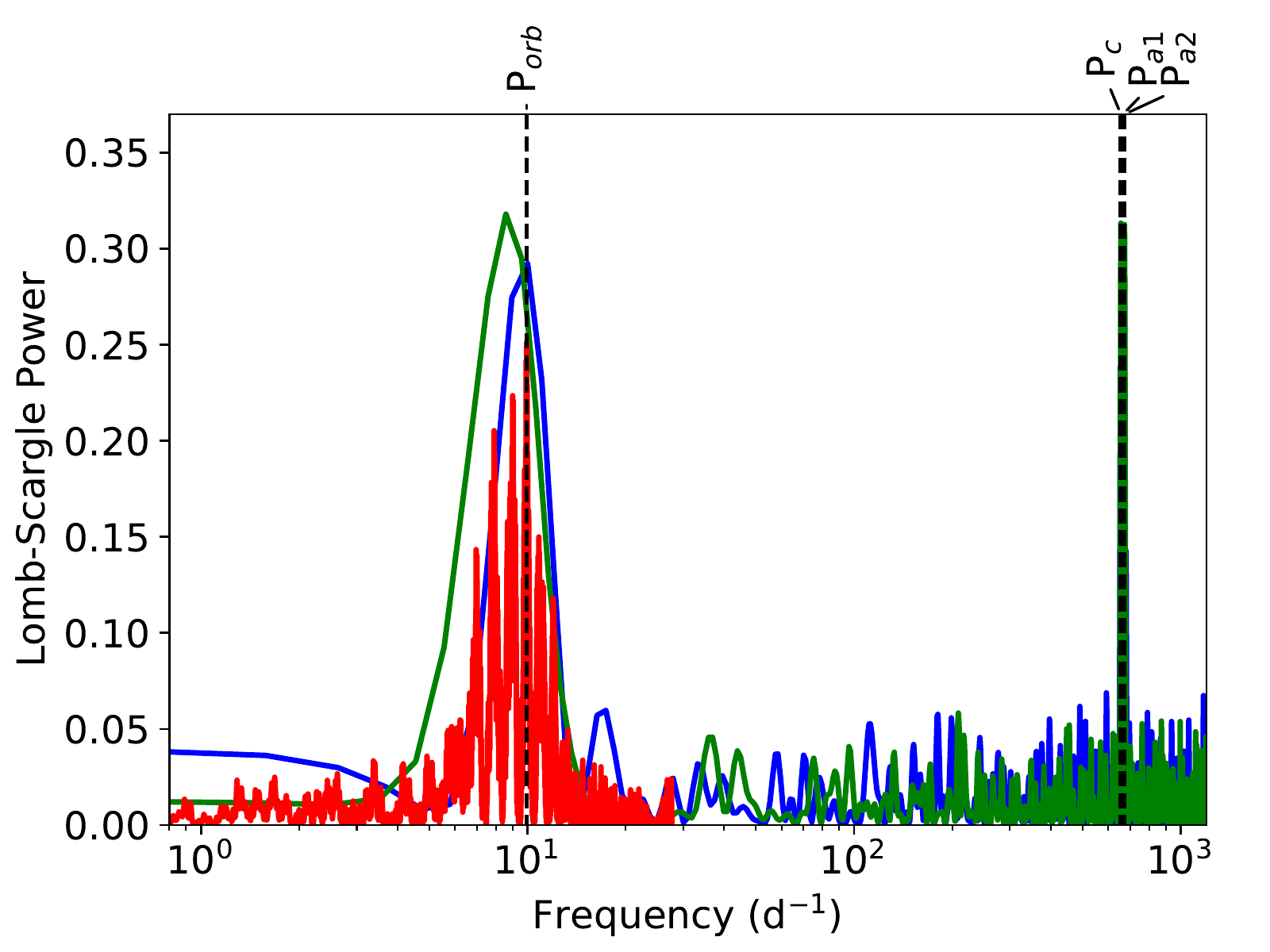}
 \caption{Lomb-Scargle periodograms of the full data set 
 (red), and the nights of 2011 May 30 (blue) and 31 (green). 
 The marked frequencies are the corresponding to the orbital period,
 ($P_{\rm orb}$), the cadence of the data ($P_c$) and once and twice 
 the actual orbital period ($P_{a1}$ and $P_{a2}$, respectively).
 }
 \label{fig:periodograms} 
 \end{figure*} 
 
Figure~\ref{fig:periodograms} shows the resulting periodogram 
of the full dataset, as well as the periodograms of the data from the 
nights of the 2011 May 30 and 31. In all instances the initial 
frequency space searched was determined by a heuristic technique 
to optimise the frequency baseline searched and the grid spacing. 
A clear peak is seen in all the periodograms at 
$\sim10\,\text{~d}^{-1}$, consistent with that reported 
by \citet{Kuulkers2013}. The periodogram of the full dataset, 
shown in red, has a complex structure due to the window function 
of the data (i.e. the gaps between the observations), whereas the 
corresponding peaks in the periodograms of the individual nights 
are broad, reflecting the fact that we are at the limit of the range 
of periods in a single night of observations. These periodograms 
also have strong peaks at $\sim$656, 666 and 673\,d$^{-1}$, 
corresponding to the beat period of the duration of the observations, 
and once and twice the actual period, respectively.
 
\begin{figure}
	\includegraphics[width=\columnwidth]{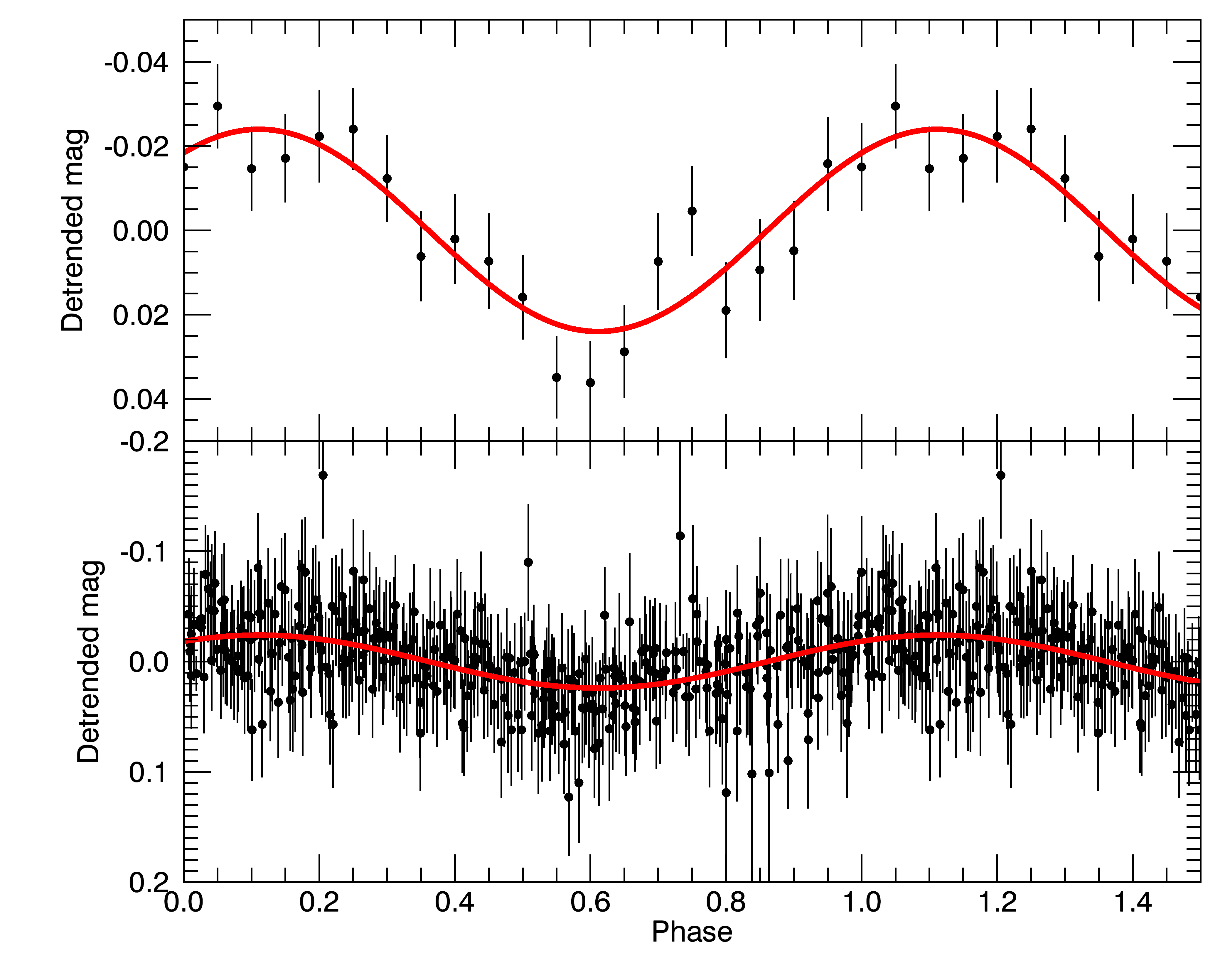}	
    \caption{\textit{Bottom:} phase folded light curve using the 
    orbital period of $2.4149\pm0.0006$\,h detected in the 
    periodogram. The phase 0 was set to the value obtained by 
    \citet{Kuulkers2013} from the central X-ray dips (MJD\,55467.0904). 
    We overplot a red sine wave calculated with the same orbital 
    period but leaving the amplitude and phase as free 
    parameters obtained through a $\chi^2$ minimization of the data. 
    The amplitude is 0.024\,mag. \textit{Top:} The same light curve 
    rebinned in 20 points.}
    \label{fig:phase}
\end{figure}

The uncertainty on the orbital period was determined via a 
bootstrap technique: 
1000 artificial light curves of \maxi were generated by sampling with 
replacement from the original dataset. The period was then estimated 
by performing a least squares fit of a simple sine wave to the data. 
Whilst we recognise that Lomb-Scargle analysis can be considered as a 
form of least squares fitting (see the review of the Lomb-Scargle 
periodgram by \citealt{VanderPlas2017}), the resulting structure of 
the window function on the periodogram makes determining the error on 
the period via Lomb-Scargle analysis difficult as we probe a discrete 
grid of frequencies. By performing a least squares fit, we can search 
a more continuous distribution of frequencies. 
The histogram of the frequencies determined from the artificial light 
curves is well described with a Gaussian with mean 9.938\,d$^{-1}$ 
and standard deviation 0.003\,d$^{-1}$. This corresponds to a period 
of 2.4149$\pm$0.0006\,h.
Figure~\ref{fig:phase} shows the complete dataset, folded on the 
determined period, along with a best fit sine wave, with frequency 
fixed to that determined by this analysis and amplitude 
free to vary. 

This is the first detection of periodic optical 
variability in MAXI\,J1659$-$152. The consistency between the X-ray period 
(i.e. $2.414 \pm 0.005$\,h) and the periodicity 
found at optical wavelengths strongly suggests that the optical variability
is due to X-ray irradiation of the donor star and not to a superhump
modulation. The latter  is due to the formation
of an eccentric and  precessing accretion disc during outburst
\citep{Whitehurst1991}. Given the orbital period, we can estimate the 
period of the superhump modulation using the empirical relation derived by 
\cite{Patterson2005}: 
\begin{equation}
\Delta P=(P_{\rm sh}-P_{\rm orb})/P_{\rm orb} = 0.18\,q + 0.29\,q^2, 
\label{equ:patt}
\end{equation}
where $P_{\rm sh}$ stands for the superhump period and $q=M_2/M_1$ for 
the mass ratio of the binary, where $M_2$ and $M_1$ are the mass of 
the secondary star and compact object, respectively. For an M5\,V  
donor star, $M_2$ ranges between 0.15 and 0.25\,M$_{\sun}$
(\citealt{Kuulkers2013}; see also section \ref{sec:discu}). On the 
other hand, \cite{Molla2016} proposed a BH mass between 4.7 
and 7.8\,M$_{\sun}$, depending on their models. These values imply 
$q=0.019-0.053$, which together with $P_{\rm orb}$, 
yield $P_{\rm sh}\sim2.422-2.439$\,h - adopting instead a canonical 
10\,M$_{\sun}$ BH, $P_{\rm sh}$ is between 2.421 and 2.425\,h for 
the range of $M_2$ mentioned above. The light curve folded with 
the 2.4149\,h period (Figure~\ref{fig:phase}) 
shows a single-humped modulation with maximum brightness occurring at
T$_0$=JD\,2455713.62643123. Assuming that this takes place at  
at orbital phase 0.5 (when the irradiated face of the donor star is
best visible to the observer), we constrain the phase of the X-ray 
dips to be $\sim0.65$.

\subsection{Study of the companion star in quiescence}
\label{sec:comp}
\cite{Kong2010} found a quiescent optical counterpart in a pre-outburst 
Pan-STARRS 
3Pi Survey image with $r\sim22.4$. However, \cite{Kennea2010} proposed 
that it was not the real counterpart since it was brighter than 
expected in the context of the \cite{Shahbaz1998} relation 
(Equation~\ref{equ:sk98}). \cite{Kong2012} reported that the 
object dropped in brightness down to $r'\sim23.7$ showing variations 
of $\sim0.2$\,mag in 30\,min interval. However, \cite{Kuulkers2013} 
discarded this counterpart and supported the line-of-sight scenario. 

\begin{figure}
	\includegraphics[width=\columnwidth]{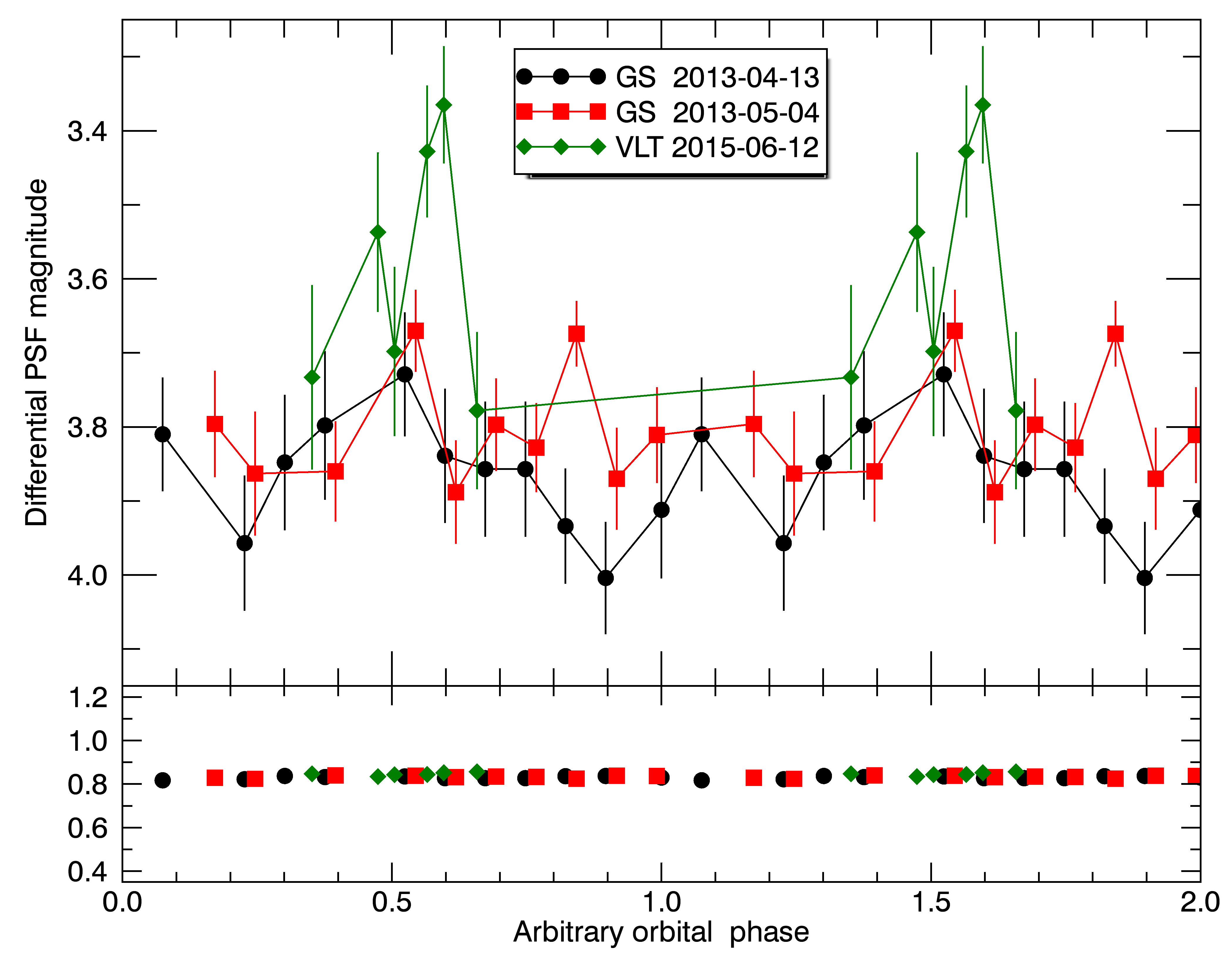}	
    \caption{Differential PSF quiescent GS and 
    VLT photometry of \maxi against C4 (top panel) as well as 
    differential photometry for the stars C4 and D8 (bottom panel).
    The data have been folded with the orbital period. The phase is 
    arbitrary. The light curve of \maxi shows no periodic trend, 
    being dominated by flickering likely produced in the accretion 
    disc.}
    \label{fig:quies}
\end{figure}

In 2014, we obtained optical photometry of MAXI\,J1659$-$152, almost 
four years after the outburst. We clearly detect the system with a 
magnitude of $24.20\pm0.08$ in the $r'$-band with GTC. This suggests 
that the system reached its true quiescent state, as shown by the trend 
in Figure~\ref{fig:evol}, between 2012 and 2014. In addition, 
we obtained images in the $i'$-band with GS in 2013 
(two nights covering $\sim2$\,h each) and $I$-band images with VLT 
in 2013 and 2015 (the latter covering 0.74\,h). On 2013 April 13 and 
May 04, we observed \maxi with GMOS-S obtaining 11 and 10 useful frames 
in each night, respectively. During the $\sim2$\,h of nightly 
coverage, the system displays optical variability with root mean 
square (RMS) of 0.2\,mag and maximum amplitude of 0.3\,mag 
(see Figure~\ref{fig:quies}).
On 2013 June 06, two images were 
acquired with FORS2 at VLT (see Torres et al. in prep.). From these 
images we obtained an average magnitude $I=23.3\pm 0.1$. In 2015, 
the system is only detected in 6 out of 11 images, with an 
average magnitude $I=23.32\pm 0.02$. The individual exposures show 
variability of up to 0.35\,mag with an RMS of 0.14\,mag, which is 
consistent with the variability shown in the GS data taken two years 
earlier (see Figure~\ref{fig:quies}).

The photometric variability observed during quiescence is consistent 
with fluctuations produced by the relative contribution of the accretion 
disc, since the time coverage by the VLT data (0.74\,h) represents 
only a third of the orbit and do not seem to follow a sinusoidal 
pattern. This activity is not surprising 
and has been reported in many others XRTs. Thus, \citet{Zurita2003}
found variability of up to 0.6\,mag independently on the orbital 
phase. However, the quality of the data does not allow us to 
establish robust conclusions about this scenario. 

We also obtained near-infrared photometry with WHT. Due 
to the variable seeing, the quiescent counterpart was only detected 
in the $H$-band images taken on 2014 July 20, with an average 
magnitude of $H=20.7\pm0.1$.

\section{Discussion}
\label{sec:discu}
The presence of an interloper star was suggested to reconcile the 
short \porb of \maxi with the amplitude of the outburst 
($\Delta V$) given by the \cite{Shahbaz1998} relation:
\begin{equation}
  \Delta V=V_{qui}-V_{out}=14.36-7.63 \log P_{orb}\,(h),
  \label{equ:sk98}
\end{equation}
where $ V_{qui} $ and $ V_{out} $ stand for the quiescent and 
outburst magnitudes in the $V$-band, respectively. Thus, for a 
2.4\,h orbital period, \cite{Kennea2011} derived a quiescent 
counterpart of $v_{qui}\sim28$ while \cite{Kuulkers2013} derived 
$V_{qui}=26.2-27.5$ (corrected by inclination effects; 
\citealt{Miller-Jones2011c}). Both authors used $V_{out}=16.5$\,mag 
at the peak of the outburst \citep{Kennea2011,Russell2010}. However, 
Equation~\ref{equ:sk98} is an empirical relation valid for systems with 
\porb$\leq12$\,h but which also may break down for \porb$\le5$\,h, 
considering that the system with the shortest orbital period found at 
that time was GRO\,J0422+32 with \porb$=5.09$\,h \citep{Gelino2003}. 
Since then, an increasing number of short orbital period XRTs have been 
discovered: XTE\,J1118+480 (\porb$=4.08$\,h; 
\citealt{Zurita2002a,Torres2004}), Swift\,J1753.5$-$0127 
(\porb$\sim3.24$\,h; \citealt{Zurita2008}) and Swift\,J1357.2$-$0933 
(\porb$=2.8$\,h; \citealt{Corral-Santana2013}). 
None of these follow Equation~\ref{equ:sk98} because they have 
quiescent magnitudes much brighter than predicted by the 
\cite{Shahbaz1998} relation.
Consequently, we suggest that the \cite{Shahbaz1998} relation breaks 
down at short orbital periods, which is specially relevant in 
MAXI\,J1659$-$152, the BH candidate with the shortest orbital period 
detected so far.

\cite{Kaur2012} found a double-peaked \ha emission line in two spectra 
obtained at the onset of the outburst. They measured 
EW(H$\alpha)=6.3\pm0.1$\,\AA, 
consistent with values reported for similar systems in outburst 
(see e.g. \citealt{Torres2002}). We obtained an 
EW(H$\alpha$)~$\sim15$\,\AA~from the colour-colour diagram during 
the decay (see Section~\ref{sec:diag}), when the system was 
$i'=22.11\pm0.04$\,mag, more than 5\,mag fainter than at the peak 
of the outburst. This difference in the EW is probably caused by 
the change in the continuum of the spectra during the evolution 
of the outburst. 
The detection of the clear \ha excess (Figure~\ref{fig:diag}) 
when the system was only 1\,mag brighter than 
in quiescence, together with the variability seen in quiescence 
(Figure~\ref{fig:quies}), clearly indicates that the counterpart proposed 
by \cite{Kong2012} is the true optical counterpart of 
MAXI\,J1659$-$152, ruling out the interloper star.

From the light curve of \maxi (Figure~\ref{fig:evol}), 
it is not possible to establish the time when the source reached 
its optical quiescent 
level. Our photometry taken with the INT on 2012 June 08 (JD\,2456086) 
about 2.5 months after \citealt{Kong2012}'s observations 
(JD\,2456009), shows that the counterpart faded to 
$r'=24.04\pm0.08$, i.e. an average rate of $\sim0.15$\,mag-per-month. 
After that, the system brightness decreased to 
$r'=24.20\pm0.08$ on 2014 June 30 (JD\,2456838). Assuming that 
the brightness of the system continued dropping constantly at that 
rate, \maxi would have reached $r'\sim24.2$ by July 2012. In addition, 
our $i'$ and $I-$band photometry, taken with GS and VLT 
(in 2013 and 2015) in three different nights, show that the system reached 
$I=23.32\pm0.02$ ($i'=23.4\pm0.1$). The average magnitude shown in 
these datasets are compatible within errors and show no trend between 
the two epochs. Thus, from the analysis of the $r'$, $I$ and 
$i'-$band magnitudes, quiescence at optical wavelengths is reached 
between July 2012 and April 2013, i.e., $>$660\,d after 
the beginning of the X-ray outburst.

On the other hand, \maxi reached its nearly quiescent X-ray 
level $\sim$220\,d after the outburst onset (see figure 4 in 
\citealt{Homan2013} and Figure~\ref{fig:evol}). 
However, a bright X-ray reflare occurred between JD\,2455700 and 
JD\,2455800 ($\sim$240\,d after the beginning of the outburst). 
\maxi returned to its quiescent X-ray level $\sim$100\,d after 
the end of the re-flare activity (i.e. $\sim$300\,d after the 
outburst was detected). This is almost a year before quiescence 
is reached at optical wavelengths. It is difficult to pin down the 
quiescence in X-rays due to the limited sensitivity of the all-sky 
monitors onboard X-ray satellites.
However, similar outburst light curve evolution was observed, e.g., 
after the 2015 June 
outburst of V404\,Cyg when the system reached the quiescent state 
in X-rays between 2015 August 5 and August 21 \citep{Sivakoff2015}, 
while in the optical the system reached this level only between 2015 
October 10 and 15 \citep{Bernardini2016b}. 
One possible explanation for the long time needed by \maxi to reach 
optical quiescence is, perhaps, that the cooling 
time of the donor 
and irradiated disc at optical wavelengths is long, while the X-rays 
coming from the central region of the accretion disc had reached 
a stable quiescent level much sooner.
This behaviour might happen in all or many black hole transients 
in outburst and with \maxi being one of the first observational 
evidences for it.\\
In order to constrain the spectral type of the companion star, 
we must analyse the relative stellar and disc contributions 
of the star and disc 
to the total light. For this, we need the unreddened flux of 
\maxi in quiescence. \cite{Kuulkers2013} reported an 
\ebv=0.26 using the estimated Galactic HI density 
($N_H=$\xe{1.74}{21}\,cm$^{-2}$) 
given by \cite{Kalberla2005} in the direction of the source. 
\cite{Kaur2012} obtained \ebv=0.32, consistent with the 
\ebv=0.34 \citep{DAvanzo2010} obtained by using 
$N_H=$\xe{2}{21}\,cm$^{-2}$ 
published by \cite{Kennea2010}, although this was later 
observed to vary with the stages of the outburst 
($N_H=$\xe{2.4-5}{21}\,cm$^{-}2$; \citealt{Kennea2011}).
This variety of reddening is not surprising since \maxi 
is a dipping source, which can often show large changes in 
their internal absorption and might lead 
to systematic errors in the distance \citep[see e.g.][]{Jonker2004}. 
Thus, for the sake of our analysis, we adopt \ebv=0.26 
\citep{Kuulkers2013}.

\begin{figure}
    \hspace*{-1.5cm}
	\includegraphics[scale=0.4]{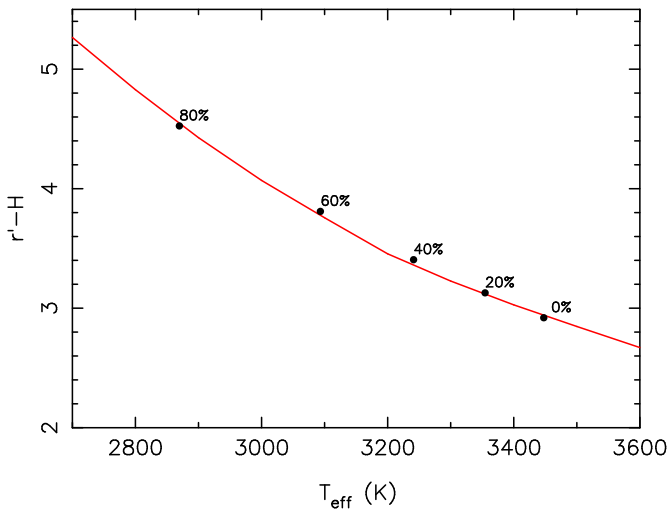}	
    \caption{($r'-H$) colour versus effective temperature 
    for NextGen model atmospheres. The red 
    line represents the model of atmospheres whereas the labelled dots are the 
    de-redderened colours of \maxi for different accretion disc 
    contaminations.}
    \label{fig:tcol}
\end{figure}

In Figure~\ref{fig:tcol} we show the the ($r'-H$) colour 
versus effective temperature for NextGen model atmospheres
\footnote{https://phoenix.ens-lyon.fr/Grids/NextGen/COLORS/}. 
We also show the de-reddened colours of \maxi allowing for the 
accretion  disc contamination to the observed flux. We first 
de-reddened the observed magnitudes of \maxi using $A_{\rm V}/E(B-V)$=3.1 
\citep{Cardelli1989}, which results in  reddening values of 
0.68 and 0.15\,mag in the $r'-$ and $H$-bands, respectively.
We then allow for different amounts of accretion disc contamination to the
observed fluxes assuming a power-law of the form $F_\lambda \propto
\lambda^{\alpha}$, as observed in X-ray binaries \citep{Shahbaz1996a}. 
The power-law index  is typically  in the range -1 to -3 and here 
we assume a value of -2.0. In Figure~\ref{fig:tcol} we show the results 
for different $r'$-band disc contaminations.
Assuming there is no disc contamination, the observed 
colours imply a companion star with $T_{\rm eff}=3450$\,K. This is an 
upper limit because, as in other X-ray systems, the observed light 
is likely contaminated by accretion disc emission. For 
short-orbital period systems such as 
GRO\,J0422+32, this could represent up to 60\% of the total light  
\citep{Filippenko1995a}. 
Assuming an index of $-2$ and  60\% $r'$-band contamination 
we find a secondary star with $T_{\rm eff}=3090$\,K.
Allowing for disc power-law indices of --1 and --3 gives  
$T_{\rm eff}=3160$\,K and $T_{\rm eff}=3070$\,K, respectively. 
Therefore, we expect the spectral type of the secondary star to lie in 
the range M2 to M5 \citep{Pecaut2013}.

\cite{Kuulkers2013} obtained $M_2=0.20$\,M$_{\sun}$~and 
$R_2=0.26$\,M$_{\sun}$ using the \cite{Smith1998} relation,
which correspond to M5\,V if the star is on 
the main-sequence. On the other hand, \cite{Jonker2012a} favour an 
M2\,V star over the M5\,V because the latter would only be consistent 
with the 4\,kpc lower limit on the distance if the accretion disc 
is dominating the optical light. 
In fact, the optical variability 
found in the VLT $I$-band images could be explained as due to a large 
contribution of the accretion disc in this band, which would favour an 
M5\,V spectral type of the companion star.
However, it is also well known that stars in cataclysmic variables and 
X-ray binaries often deviate from main-sequence and tend to be 
overluminous for their masses and colors, which would imply a hotter 
spectral type.

\section{Conclusions}
\label{sec:concl}
We present optical and infrared data of the evolution of \maxi 
during its 2010 outburst, decay and quiescence, analysing more than 400 
images covering 5 years of photometric data. 
\begin{itemize}
\item[(i)] Combining several 
observational campaigns of \maxi to 
monitor the decay towards quiescence 
we obtained the first optical confirmation of the $2.414\pm0.005$\,h 
orbital period.
This value is in perfect agreement with that  
obtained by \cite{Kuulkers2013} analysing the X-ray dips, confirming 
\maxi as the BH XRT with the shortest orbital period hitherto. 
The light curve folded with the orbital 
period is single hump and likely produced by X-ray irradiation 
of the donor star. 
Assuming that the maximum of the light curve occurs at 
phase 0.5, the outburst X-ray dips would have occurred  
at phase 0.65.\\

\item[(ii)] In addition, we investigated a \diag diagram 
with all the stars in the field. In this diagram, the optical 
counterpart of \maxi during its outburst decay is clearly detected above the 
main stellar locus, implying a large \ha excess.\\

\item[(iii)] We also determined optical quiescent magnitudes of MAXI\,J1659$-$152: 
$r'=24.20\pm0.08$, $I=23.32\pm0.02$ and $H=20.7\pm0.1$. 
Combining the $r'-H$ colour together with a model of stellar atmospheres 
and assuming a 60\% accretion disc contamination to the observed $r'$-band 
fluxes, we find a secondary star in the range M2 to M5.
This range of spectral types are in agreement with those reported by 
\cite{Kuulkers2013} and \cite{Jonker2012a}.
\end{itemize}
\section*{Acknowledgements}

We thank the anonymous referee for the useful comments. 
JMC-S acknowledges partial support from CONICYT-Chile through the FONDECYT 
Postdoctoral Fellowship 3140310. MAPT acknowledges support via a 
Ram\'on y Cajal Fellowship (RYC-2015-17854). TS acknowledges support 
by the Spanish Ministry of Economy and Competitiveness (MINECO) 
under the grant AYA2013-42627. JC acknowledges support by the 
Leverhulme Trust through the Visiting Professorship grant VP2-2015-04. 
TMD acknowledges support via a Ram\'on y Cajal Fellowship (RYC-2015-18148)
FEB acknowledges support from CONICYT-Chile grants Basal-CATA PFB-06/2007, 
FONDECYT Regular 1141218 and the Ministry of Economy, Development and 
Tourism's Millenium Science Initiative through grant IC120009, awarded to 
The Millenium Institute of Astrophysics, MAS. PGJ acknowledges support 
from the European Research Council (ERC CoG-647208).
\\
This article is based on observations made with several telescopes 
installed in the Spanish Observatorio del Roque de los Muchachos of 
the Instituto de Astrof\'isica de Canarias, in the island of La Palma.
MER is operated by the Flemish Community whereas the INT, WHT and their 
service programme are operated by the Isaac Newton Group of telescopes. 
The Faulkes Telescope Project is an education partner of Las Cumbres 
Observatory. The Faulkes Telescopes are maintained and operated by LCO. 
Based on observations 
collected at the European Organisation for Astronomical Research in the 
Southern Hemisphere under ESO programmes 091.D-0865(A) and 095.D-0973(A). 
Based on observations obtained at the Gemini Observatory (under program ID 
GS-2013A-Q-58), which is operated by the Association of Universities for 
Research in Astronomy, Inc., under a cooperative agreement with the 
National Science Foundation (NSF) on behalf of the Gemini partnership: the 
NSF (United States), the National Research Council (Canada), Comisi\'on 
Nacional de Investigaci\'on Cient\'ifica y Tecnol\'ogica (Chile), 
Ministerio de Ciencia, Tecnolog\'{i}a e Innovaci\'{o}n Productiva 
(Argentina), and Minist\'{e}rio da Ci\^{e}ncia, Tecnologia e 
Inova\c{c}\~{a}o (Brazil).\\

%
%




\bibliographystyle{mnras}
\bibliography{biblio} 




%
%


\bsp	
\label{lastpage}
\end{document}